\documentclass[11pt]{article}
\usepackage{amsmath,amsthm,amsfonts,amssymb}

\newcommand{\hilb}{\mbox{Hilb}}
\newcommand{\cycl}{\mbox{Cycl}}

\newtheorem{teo}{Theorem}[section]
\newtheorem{coro}[teo]{Corollary}
\newtheorem{prop}[teo]{Proposition}

\newtheorem{obs}[teo]{Remark}

\begin{document}

\title{{Some remarks on the GNS representations of topological $^*$-algebras}}

\author{S.~M.~Iguri\footnote{e-mail: siguri@iafe.uba.ar} \, and M.~A.~Castagnino\footnote{e-mail: oningatsac@iafe.uba.ar}}

\date{\small Instituto de Astronom\'{\i}a y F\'{\i}sica del Espacio (CONICET-UBA).\\ C.~C.~67 - Suc.~28, 1428 Buenos Aires, Argentina. \\ and \\ Dpto. de F\'\i sica, FCEyN, Universidad de Buenos Aires. \\ Ciudad Universitaria Pab. I, 1428 Buenos Aires, Argentina.}

\maketitle

\begin{abstract}
After an appropriate restatement of the GNS construction for topological $^*$-algebras we prove that there exists an isomorphism among the set $\cycl(A)$ of weakly continuous strongly cyclic $^*$-representations of a barreled dual-separable $^*$-algebra with unit $A$, the space $\hilb_A(A^*)$ of the Hilbert spaces that are continuously embedded in $A^*$ and are $^*$-invariant under the dual left regular action of $A$ and the set of the corresponding reproducing kernels. We show that these isomorphisms are cone morphisms and we prove many interesting results that follow from this fact. We discuss how these results can be used to describe cyclic representations on more general inner product spaces.
\end{abstract}

{\small \noindent 2000 Mathematics Subject Classification : 16G99, 46H15, 47L90, 81P99, 81R15.

}

\section {Introduction}

Quantum statistical mechanics and quantum field theories are believed to be fully described in purely algebraic terms, the so-called C$^*$-algebraic approach (see \cite{bratelli,bratelli2,haag} for textbooks and \cite{buchholz,borchers} for recent reviews on the subject) being the most appealing one. Despite of the successful aspects of the C$^*$-algebraic approach, in order to find abstract counterparts for all observable magnitudes in an algebraic approach it is mandatory to consider $^*$-algebras with less restrictive topologies than the ones derived from C$^*$-norms \cite{iguri2,bagarello,inoue,voronin}. Moreover, if quantum gauge theories are also assumed to be described in algebraic terms, the appropriate representation spaces would be more general inner product spaces than Hilbert spaces \cite{albeverio} and in that case there is no compelling reason to believe that the $^*$-algebra describing the observable content of the theory should be a normable one. 

One of the fundamentals of the C$^*$-algebraic approach is the Gelfand-Naimark-Segal (GNS) theorem. The so-called GNS construction is an important tool from both the physical and the strictly mathematical points of view. It characterizes the building blocks of the representation theory of C$^*$-algebras, i.e., their cyclic representations and defines in this way the bridge between the formalism and the physical reality.

During the 70's the systematic study of the representations of algebras of unbounded operators begun with Powers \cite{powers,powers2,gudder}. In the seminal paper of Powers there is a version of the GNS theorem but unfortunately it makes no mention on the topological properties of the represented $^*$-algebra. The lack of information on this topology gives to the construction generality but, on the other hand, it restricts its scope. There were other statements of the GNS theorem during the last years \cite{palmer1,palmer2,iguri1,stochel} assuming more or less restrictive conditions on the topological nature of the $^*$-algebra and there are even versions of the GNS construction on non-necessarily definite positive inner product spaces \cite{antoine,hofmann,mnatsakanova}. The aim of this paper is to complement all these treatments.

We will restate the GNS theorem for a wide class of topological $^*$-algebras this restatement allowing us to prove that there exists a continuous bijection between the space of GNS representations and a set of Hilbert spaces continuously embedded in the dual space of the $^*$-algebra in hands, an idea already suggested in \cite{belanger,constantinescu}. More explicitly, if $A$ is a barreled dual-separable $^*$-algebra with unit we will prove that the set $\cycl(A)$ of weakly continuous strongly cyclic $^*$-representations of $A$ is isomorphic to the set $\hilb_A(A^*)$ of the Hilbert subspaces of $A^*$ that are $^*$-invariant under the left dual regular action of $A$ on $A^*$. In turn this bijection can be extended to a multiple isomorphism among these spaces, the space of continuous positive functionals over $A$ and the corresponding space of invariant positive operators.

This characterization of the space of GNS representations will also allow us to transport the cone structure already defined on $\hilb_A(A^*)$ to $\cycl(A)$ and to prove some remarkable consequences of this fact. Let us mention that this strictly convex cone structure on $\cycl(A)$ can be used for describing GNS representations over spaces with non-necessarily positive definite inner product \cite{casta} and it could be useful for studying deformation theory of GNS representations \cite{waldman}.

The paper is organized as follows. In section \ref{hilbertsubsp} we review the main results of Schwartz's theory of Hilbert subspaces \cite{schwartz} and their associated reproducing kernels, the most important one being the natural bijection between the set of Hilbert subspaces of a given topological space and the set of positive operators mapping its dual on it. In section \ref{representation} we present those aspects of the representation theory of topological algebras needed in the sequel. We have essentially followed \cite{schmudgen} but some concepts were slightly modified. In section \ref{gnshilb} we show that for a barreled dual-separable $^*$-agebra there is a one-to-one correspondence between its GNS representations and those Hilbert subspaces of its dual that are invariant under the dual left regular action. We also show that this map is a cone morphism for the cone structure already defined on this last space in \cite{schwartz}. In section \ref{consequences} we derive many consequences of the previous sections. Finally, in section \ref{concluding} we present our conclusions.

\section{Hilbert subspaces and reproducing kernels}
\label{hilbertsubsp}

In this section we will review some definitions and we will introduce a few items of notation concerning the theory of the Hilbert spaces that can be continuously embedded in a quasi-complete\footnote{A topological vector space $E$ is said to be {\it quasi-complete} if every bounded closed subset of $E$ is complete \cite{treves}.} locally convex Hausdorff separable vector space over the field of complex numbers ${\mathbb{C}}$ \cite{schwartz}. We will denote any space fulfilling these requirements by $E$.

Let us first recall the definition of a Hilbert subspace of $E$. A linear subspace $\mathcal{H} \subseteq E$ is called a {\it Hilbert subspace} of $E$ whenever $\mathcal{H}$ is equipped with a definite positive inner product $(\cdot,\cdot)$ turning it into a Hilbert space and the inclusion of $\mathcal{H}$ into $E$ is a continuous map, the norm $\|\cdot\|=(\cdot,\cdot)^{1/2}$ defining a topology on $\mathcal{H}$ finer than the one induced by $E$.

When dealing with the Hilbert subspaces of $E$ it is convenient to consider the topological anti-dual space of $E$ instead of its dual $E'$, the reason being that every Hilbert space can be canonically identified through the Riesz isomorphism with its anti-dual. The {\it anti-dual space} $E^*$ of $E$ is the conjugate of $E'$\footnote{Notice that, up to isomorphisms, $E^*$ is unique.}, i.e., it is defined as a topological vector space over ${\mathbb{C}}$ with an anti-isomorphism mapping it onto $E'$. Under this map the canonical bilinear form on $E' \times E$ becomes a sesquilinear map on $E^* \times E$ which we will denote as $\left\langle x | \phi \right\rangle$ for all $x \in E^*$ and all $\phi \in E$. Notice that this bracket is anti-linear in its first argument while it is linear in the second one. We will refer to the elements of $E^*$ as {\it functionals} over $E$ even when it should be remembered that they are not elements of the dual. If $E$ is a Hilbert space and, as we have already said, we identify the elements of $E$ with those of $E^*$, the duality bracket reduces to the inner product on $E$.

As in the case for a strictly dual system, given a continuous map $T:E \rightarrow F$ we introduce its {\it adjoint} as the linear map $T^*:F^* \rightarrow E^*$ defined by the identity $\left\langle T^*x | \phi \right\rangle = \left\langle x | T\phi \right\rangle$ for all $x \in F^*$ and all $\phi \in E$. It is a continuous map provided that $E^*$ and $F^*$ are equipped with their weak or, alternatively, their strong dual topologies. If both spaces are Hilbert spaces, the adjoint of a map equals the usual Hilbert space adjoint, i.e., $T^*=T^{\dag}$. In the case that $T$ is an anti-linear operator, the expression defining its adjoint must be replaced by $\left\langle T^*x | \phi \right\rangle = \overline{\left\langle x | T\phi \right\rangle}$ for all $x \in F^*$ and all $\phi \in E$.

Let $\mathcal{H}$ be a Hilbert subspace of $E$ and let $J$ be the inclusion map of $\mathcal{H}$ into $E$. By the Riesz representation theorem, given $x\in E^*$, there exists a unique element $J^* x \in \mathcal{H}$ such that $\left( J^* x , \xi \right) = \left\langle x | J \xi \right\rangle$ for all $\xi \in \mathcal{H}$. Let us denote by $Hx = JJ^*x$ the same element regarded as an element of $E$. The operator $H$ mapping $E^*$ into $E$ is called the {\it reproducing operator} of $\mathcal{H}$. It is a continuous operator when $E$ and $E^*$ are equipped with their weak topologies $\sigma(E^*,E)$ and $\sigma(E,E^*)$, respectively.

The reproducing operators of Hilbert subspaces have many remarkable properties. For instance, they are all {\it hermitian}, where by a hermitian operator we mean a linear map $T:E^* \rightarrow E$ satisfying $\left\langle x|Ty \right\rangle = \overline{\left\langle y|Tx \right\rangle}$, for any pair $x,y \in E^*$. In fact, for all $x,y \in E^*$ we have $\left\langle x|Hy \right\rangle = \left( J^* x, J^* y \right) = \overline{\left( J^* y, J^* x \right)}= \overline{\left\langle y|Hx \right\rangle}$. Moreover, setting $x = y$ in the last expression it follows that $H$ is a {\it positive operator}, i.e., $\left\langle x|Hx \right\rangle = \left( J^* x,J^* x \right) \geq 0$ for all $x \in E^*$. In addition, it is possible to prove a Cauchy-Schwartz like identity, i.e., $\left|\left\langle x|Hy \right\rangle\right|^2 \leq \left\langle x|Hx \right\rangle \left\langle y|Hy \right\rangle$ for all $x,y \in E^*$.

Let us denote by $L(E)$ the set of all continuous operators mapping $E^*$ into $E$, these spaces being endowed with their weak topologies, and let $L^+(E)$ be the proper strictly convex cone of positive elements of $L(E)$. Reproducing operators belong to $L^+(E)$.

The map that assigns to each element $T \in L(E)$ the form given by $\left\langle x|Ty\right\rangle$ with $x,y \in E^*$ is an algebraic isomorphism mapping $L(E)$ onto the space of separately weakly continuous sesquilinear forms on $E^*$, i.e., the {\it kernels} on $E^*$. As it should be clear, when this map is restricted to $L^+(E)$ it gives an isomorphism onto the {\it positive kernels} on $E^*$. In this context, if the operator in $L^+(E)$ is the reproducing operator of a Hilbert subspace $\mathcal{H}$ of $E$, the corresponding sesquilinear form on $E^*$ is called the {\it reproducing kernel} of $\mathcal{H}$ in $E$.

Let $\hilb(E)$ be the set of all Hilbert subspaces of $E$. As it was already proved by Schwartz \cite{schwartz}, it is possible to endow $\hilb(E)$ with a proper strictly convex cone structure. Let us briefly outline the corresponding definitions.

{\bf Sum of Hilbert subspaces:} Let $\mathcal{H}_1$ and $\mathcal{H}_2$ be two Hilbert subspaces of $E$, $J_1$ and $J_2$ being the respective inclusion maps. Let $\mathcal{H}_1 \times \mathcal{H}_2$ be the Hilbert space product of $\mathcal{H}_1$ and $\mathcal{H}_2$. Finally, let $\Phi:\mathcal{H}_1 \times \mathcal{H}_2 \rightarrow E$ be the continuous map given by $\Phi(\xi_1,\xi_2)=J_1\xi_1+J_2\xi_2$ and consider the quotient space $(\mathcal{H}_1 \times \mathcal{H}_2)/\ker(\Phi)$ equipped with its canonical Hilbert space structure. The {\it sum of $\mathcal{H}_1$ and $\mathcal{H}_2$} is defined as the image space $\Phi(\mathcal{H}_1 \times \mathcal{H}_2) \subseteq E$ endowed with the unique norm that makes the canonical linear bijection between $(\mathcal{H}_1 \times \mathcal{H}_2)/\ker(\Phi)$ and $\Phi(\mathcal{H}_1 \times \mathcal{H}_2)$ an isometric isomorphism. We will denote this space by $\mathcal{H}_1+\mathcal{H}_2$. The norm on $\mathcal{H}_1+\mathcal{H}_2$ is explicitly given by $\|\xi\|^2=\inf \{ \|\xi_1\|_1^2+\|\xi_2\|_2^2 \}$ where $\|\cdot\|_1$ (resp., $\|\cdot\|_2$) is the norm on $\mathcal{H}_1$ (resp., $\mathcal{H}_2$) and the infimum is taken over those pairs $(\xi_1,\xi_2)\in \mathcal{H}_1 \times \mathcal{H}_2$ such that $\xi=\Phi(\xi_1,\xi_2)$. If $\ker(\Phi)=0$, then $\mathcal{H}_1 \cap \mathcal{H}_2 = \{ 0 \}$ and $\mathcal{H}_1+\mathcal{H}_2$ is simply the Hilbert space direct sum of both spaces and in that case we will write, as it is usual, $\mathcal{H}_1 \oplus \mathcal{H}_2$.

The definition of the sum of two Hilbert subspaces does agree with a more general construction concerning the Hilbert subspaces of spaces that are images under continuous mappings. Let $E$ and $F$ be two quasi-complete locally convex Hausdorff separable vector spaces over ${\mathbb{C}}$ and let $T:E \rightarrow F$ be a continuous linear map. Consider a Hilbert subspace $\mathcal{H}$ of $E$ and let us denote by $J$ the inclusion map of $\mathcal{H}$ into $E$. Since $TJ: \mathcal{H} \rightarrow F$ is also a continuous map, its kernel is a closed linear subspace of $\mathcal{H}$. The image space of $\mathcal{H}$ under $TJ$, endowed with the Hilbert structure making the restriction of $TJ$ to $\ker (TJ)^{\bot}$ a linear isometry, is a Hilbert subspace of $F$. We will simply denote this space by $T(\mathcal{H})$.

{\bf Multiplication by non-negative real numbers in $\hilb(E)$:} The multiplication law on $\hilb(E)$ by non-negative real numbers is defined as follows. Let $\mathcal{H}$ be a Hilbert subspace of $E$ and let $\lambda$ be a positive real number. The space $\lambda \mathcal{H}$ is the Hilbert subspace of $E$ with underlying linear space equal to $\mathcal{H}$ and the norm on $\lambda \mathcal{H}$ being defined as $(1/\sqrt{\lambda})$ times the original norm on $\mathcal{H}$. The action of $\mathbb{R}_{>0}$ on $\hilb(E)$ is extended to $\mathbb{R}_{\ge 0}$ setting $\lambda \mathcal{H}$ equal to $\{0\}$ when $\lambda=0$.

{\bf Order in $\hilb(E)$:} Finally, a partial order, compatible with the structures given above, is defined on $\hilb(E)$ in the following way. If $\mathcal{H}_1$ and $\mathcal{H}_2$ are two Hilbert subspaces of $E$, we will write $\mathcal{H}_1 \leq \mathcal{H}_2$ if $\mathcal{H}_1 \subseteq \mathcal{H}_2$ and the inclusion of $\mathcal{H}_1$ into $\mathcal{H}_2$ is an operator of norm at most 1, i.e., $\mathcal{H}_1$ belongs to $\hilb(\mathcal{H}_2)$.

If $\mathcal{H}_1$ and $\mathcal{H}_2$ are two Hilbert subspaces of $E$ it is easy to check that $\mathcal{H}_1 \cap \mathcal{H}_2 = \{0\}$ if and only if $\mathcal{H}_1$ and $\mathcal{H}_2$ are {\it mutually excluding} for the order relation in $\hilb(E)$, i.e., if for any Hilbert subspace $\mathcal{K}$ of $E$ such that $\mathcal{K} \le \mathcal{H}_1$ and $\mathcal{K} \le \mathcal{H}_2$ it follows that $\mathcal{K} =\{0\}$. We will say that a Hilbert subspace $\mathcal{H}$ is {\it indecomposable} if it does not admit a non-trivial decomposition as a direct sum of Hilbert subspaces, i.e., if for any decomposition $\mathcal{H}=\mathcal{H}_1 + \mathcal{H}_2$, $\mathcal{H}_1$ and $\mathcal{H}_2$ being two mutually excluding Hilbert subspaces of $E$, it is possible to prove that $\mathcal{H}_1=\{0\}$ or $\mathcal{H}_2=\{0\}$. We will denote by $[0,\mathcal{H}]$ the interval in $\hilb(E)$ between $\{0\}$ and $\mathcal{H}$, i.e., $[0,\mathcal{H}]=\{\mathcal{H}'\in \hilb(E) : 0 \le \mathcal{H}' \le \mathcal{H}\}$. We will say that $\mathcal{H}$ is an {\it extremal element} of $\hilb(E)$ if the interval $[0,\mathcal{H}]$ equals $\{\lambda \mathcal{H} \in \hilb(E):0\le\lambda\le 1\}$. Notice that every extremal element of $\hilb(E)$ is an indecomposable Hilbert subspace of $E$ but the converse of this statement is not generally true.

We are now in a position to recall the most important result of Schwartz Hilbert subspaces theory. We will only sketch the proof and we refer to \cite{schwartz} for more details.

\begin{teo}
\label{teo1}
The map that assigns to each Hilbert subspace of $E$ its reproducing operator is a bijection from $\hilb(E)$ onto $L^+(E)$.
\end{teo}

\begin{proof}
Let $\mathcal{H}$ be a Hilbert subspace of $E$ and let us denote by $J$ the inclusion of $\mathcal{H}$ in $E$. We will prove that $\mathcal{H}$ is determined by its reproducing operator $H$. First, notice that $J$ being an inclusion, it follows that $J^*$ is a dense range projection, i.e., $J^* E^*$ is a dense linear subspace of $\mathcal{H}$. On the other hand, there is no element in $\mathcal{H}$ orthogonal to $J^* E^*$ but the origin. Since the unit ball $B$ of $\mathcal{H}$ is weakly compact, $JB$ is weakly closed in $E$, and this set being convex, it is closed for the original topology on $E$. It follows that $JB$ is the closure in $E$ of the set $\{ Hx\in E:\left\langle x|Hx\right\rangle^{1/2}\leq 1 \}$ and this proves that $\mathcal{H}$ is fully determined by $H$. Moreover, given $\phi \in E$, it belongs to $\mathcal{H}$ if and only if $\sup \{ \left|\left\langle x| \phi \right\rangle \right| / \left\langle x|Hx\right\rangle^{1/2} \} < +\infty$, where the supremum is taken over the elements $x\in E$ such that $\left\langle x|Hx\right\rangle > 0$.
The value of this expression equals $\|\phi\|$.

Now, let us prove that given $H \in L^+(E)$, we can define a Hilbert subspace $\mathcal{H}$ of $E$ in such a way that its reproducing operator equals it. Let the quotient space $E^*/\ker(H)$ be equipped with the Hilbert space structure derived from the sesquilinear form induced by $H$. The canonical injection of $E^*/\ker(H)$ into $E$ has a one-to-one extension to the completion. It is in order to prove this statement that it is essential to assume that $E$ is a quasi-complete separable vector space \cite{schwartz}. The image space under this mapping, endowed with the unique Hilbert space structure turning it into an isometry, is a Hilbert subspace of $E$, its reproducing operator being $H$.
\end{proof}

\begin{prop}
\label{prop1}
The bijection between $L^+(E)$ and $\hilb(E)$ is a cone morphism for the usual cone structure on $L^+(E)$ and the cone structure we have already introduced for $\hilb(E)$. More explicitly, let $\mathcal{H}$, $\mathcal{H}_1$ and $\mathcal{H}_2$ be three Hilbert subspaces of $E$, let $H$, $H_1$ and $H_2$ be their respective reproducing operators. Let $\lambda$ be a non-negative real number. Then:
\begin{enumerate}
\item[1. ]$\mathcal{H}=\mathcal{H}_1+\mathcal{H}_2$ if and only if $H=H_1+H_2$,
\item[2. ]$\mathcal{H}=\lambda \mathcal{H}_1$ if and only if $H= \lambda H_1$, and
\item[3. ]$\mathcal{H}_1\le \mathcal{H}_2$ if and only if $H_1\le H_2$.
\end{enumerate}
\end{prop}

\begin{proof}
See \cite{schwartz}, Prop. 11-13, p. 158-160.
\end{proof}

\section{Representations of $^*$-algebras}
\label{representation}

The purpose of this section is to restate the classical GNS theorem for topological $^*$-algebras. In order to do that we will first recall some basic facts on representations and $^*$-representations of $^*$-algebras on Hilbert spaces by (non necessarily bounded) linear operators. We will mainly follow \cite{schmudgen} except for some minor changes in some definitions that will be justified in the rest of the paper. We will always denote by $A$ an associative algebra over ${\mathbb{C}}$. When assuming $A$ to be an algebra with unit, we will denote the unit in $A$ by $e$.

A {\it representation $\pi$ of $A$} on a Hilbert space $\mathcal{H}$ is a map from $A$ into a set of linear operators, all of them defined on a common domain $D$, such that the following conditions are fulfilled:
\begin{enumerate}
\item[1.] $D$ is a dense subspace of $\mathcal{H}$,
\item[2.] $D$ is invariant under the action of $A$, i.e., $\pi(x)D\subseteq D$ for all $x\in A$, and
\item[3.] $A$ acts linearly and multiplicatively on $D$, i.e., for all $x,y \in A$ and all $\lambda \in \mathbb{C}$ we have: $\pi(x+y)=\pi(x)+\pi(y)$, $\pi(\lambda x)=\lambda \pi(x)$ and $\pi(xy)=\pi(x)\pi(y)$.
\end{enumerate}
If the algebra has a unit, it is also assumed that
\begin{enumerate}
\item[4.] $\pi(e)$ equals the identity operator on $D$, i.e., $\pi(e)=\mbox{Id}_{D}$.
\end{enumerate}

Let $\pi_1$ and $\pi_2$ be two representations of $A$ on the Hilbert spaces $\mathcal{H}_1$ and $\mathcal{H}_2$ and let $D_1$ and $D_2$ be their respective domains. We will say that $\pi_1$ is an {\it algebraic extension} of $\pi_2$ or that $\pi_2$ is an {\it algebraic subrepresentation} of $\pi_1$, and we will write $\pi_2 \subseteq \pi_1$, if $D_2 \subseteq D_1$, $\mathcal{H}_2$ is a linear subspace of $\mathcal{H}_1$ and $\pi_1(x) \upharpoonright D_2$ equals $\pi_2(x)$ for every $x\in A$. If, in addition, the scalar product on $\mathcal{H}_2$ is the restriction to $\mathcal{H}_2$ of the scalar product on $\mathcal{H}_1$, i.e., $\pi_2(x) \subseteq \pi_1(x)$ for all $x \in A$, we will say that $\pi_1$ is an {\it extension} of $\pi_2$ or that $\pi_2$ is a {\it subrepresentation} of $\pi_1$. In this last case we will write $\pi_2 \le \pi_1$.

\begin{obs} \label{obs} In the next section we will find that the conditions imposed on $\mathcal{H}_1$ and $\mathcal{H}_2$ for $\pi_2$ to be a subrepresentation of $\pi_1$ can be conveniently modified. We will consider a less restrictive notion of extension of a representation asking $\mathcal{H}_2$ to be a Hilbert subspace of $\mathcal{H}_1$, i.e., we will assume that $\mathcal{H}_2$ is continuously embedded in $\mathcal{H}_1$, the corresponding inclusion being an operator of norm at most $1$. Let us notice that all the contents of the present section will remain being valid. \end{obs}

In order to define a concept analogous to the one of a closed operator but for a representation, we will proceed as usual endowing the domain of a given representation with a topology induced by the action of the algebra on it. Let $\pi$ be a representation of $A$ on a Hilbert space $\mathcal{H}$ and let $D$ be its domain. The {\it graph topology} on $D$ is the locally convex topology generated by the family of seminorms $\{p_x = \left\| \pi(x) \, \cdot \right\|\}$ where $\left\| \, \cdot \, \right\|$ is the norm on $\mathcal{H}$ and $x$ runs over $A$. The graph topology can be characterized as the weakest locally convex topology on $D$ which makes the embedding of $D$ into its completion relative to the topology determined by the norm $\left\| \, \cdot \, \right\| + \left\| \, \pi(x) \cdot \, \right\|$ a continuous mapping for every $x \in A$. In this context, the graph topology can be viewed as a projective topology in the sense of the theory of locally convex spaces \cite{treves}. When $A$ has a unit, the graph topology is always finer than the one induced by $\mathcal{H}$ on $D$. Clearly, the graph topology is generated by a single norm, the one on $\mathcal{H}$, if and only if the image of each element of $A$ through $\pi$ can be extended to a bounded operator on $\mathcal{H}$. 

If $D$ is complete when equipped with the graph topology we will say that $\pi$ is a {\it closed representation}. A representation $\pi$ will be called a {\it closable} representation of $A$ if $\pi(x)$ is a closable operator on $D$ for all $x \in A$.

Given a closable representation of $A$ on a Hilbert space $\mathcal{H}$, let $\overline{\pi(x)}$ be the closure of the operator $\pi(x)$, the domain of $\pi(x)$ being the common domain $D$ for all $x \in A$. Let us denote by $\overline{D}_x$ the domain of $\overline{\pi(x)}$ for every $x\in A$. Finally, let $\overline{D}$ be the completion of $D$ in $\bigcap_{x \in A} \overline{D}_x$ relative to the graph topology. $\overline{D}$ is the domain of a closed representation $\overline{\pi}$ of $A$ in $\mathcal{H}$ defined as
\begin{equation}
\overline{\pi}(x) = \overline{\pi(x)} \upharpoonright \overline{D}
\end{equation}
for all $x \in A$. This representation is called the {\it closure} of $\pi$, and it is the minimal closed extension of it. Of course \cite{powers} $\pi$ is closed if and only if $\pi$ is closable and it equals $\overline{\pi}$.

Let us now assume that $A$ is a $^\star$-algebra. Like in the case of a single operator acting on a Hilbert space, we will define an adjoint of a given representation.

Suppose that $\pi$ is a representation of $A$ on a Hilbert space $\mathcal{H}$ and let $D$ be its domain. For all $x \in A$, let $\pi(x)^*$ be the adjoint of $\pi(x)$ and let $D^*_x$ be its domain. Further, let $D^*=\bigcap_{x \in A} D^*_x$ and let us denote by $\mathcal{H}^*$ the completion of $D^*$ in $\mathcal{H}$. The {\it adjoint representation} of $\pi$ is defined as the representation $\pi^*$ of $A$ on $\mathcal{H}^*$ with domain $D^*$ given by
\begin{equation}
\pi^*(x)=\pi(x^*)^* \upharpoonright D^*
\end{equation}
for all $x\in A$.

The adjoint of a given representation is always a closed representation and it is the largest one among those representations $\tilde{\pi}$ of $A$ on $\mathcal{H}^*$ with domain $\tilde{D}$ that satisfies $\left(\xi,\tilde{\pi}(x^*) \chi\right)=\left(\pi(x) \xi, \chi \right)$ for all $x \in A$, $\chi \in \tilde{D}$ and $\xi \in D$.

We will say that a representation $\pi$ is {\it adjointable} (resp. {\it biclosed}) if $\mathcal{H}=\mathcal{H}^*$ (resp. if it equals its biadjoint representation, i.e., if $\pi^{\star \star} = (\pi^*)^*$).

All concepts above suggest the following definition originally introduced by Powers in \cite{powers}. Let $\pi$ be a representation of a $^*$-algebra $A$ on a Hilbert space $\mathcal{H}$ and let $D$ be its dense domain. We will say that $\pi$ is a {\it hermitian representation}, or simply a {\it $^*$-representation of $A$}, if $\pi \le \pi^*$, in other words, if for all $\chi, \xi \in D$ and all $x \in A$, $\pi$ satisfies
\begin{equation}
\left( \xi, \pi(x^*) \chi \right) = \left( \pi(x) \xi ,\chi\right)
\end{equation}

Notice that every $^*$-representation is necessarily adjointable.

If $A$ is a Banach $^*$-algebra then a $^*$-representation $\pi$ of $A$ on a Hilbert space $\mathcal{H}$ is closed if and only if $D$ equals $\mathcal{H}$. On the other hand, if $\pi$ is a $^*$-representation of a $^*$-algebra $A$ on a Hilbert space $\mathcal{H}$ and $D = \mathcal{H}$, it follows from the closed graph theorem that $\pi$ is a bounded representation, i.e., $\pi$ maps $A$ into bounded operators on $\mathcal{H}$. These facts are clear evidences that the previous definition is a consistent generalization of the usual concept of $^*$-representation by bounded operators.

The adjoint representation of a given $^*$-representation $\pi$ may fail to be a $^*$-representation as it is the case of the adjoint of a single hermitian operator acting on a Hilbert space. But as we have already said it is actually a closed representation extending $\pi$. Moreover, every $^*$-representation extending $\pi$ is necessarily a restriction of $\pi^*$.

We will say that a $^*$-representation $\pi$ of a $^*$-algebra $A$ is a {\it maximal} (resp. {\it self-adjoint}, resp. {\it essentially self-adjoint}) $^*$-representation if every $^*$-representation extending $\pi$ equals it (resp. if $\pi=\pi^*$, resp. if its closure is a self-adjoint representation.)

Some general properties of $^*$-representations are collected in the following proposition. The proof can be found in \cite{schmudgen}.

\begin{prop}
\label{prop2}
Let $\pi$ be a $^*$-representation of $A$ in a Hilbert space $\mathcal{H}$ with domain $D$.
\begin{enumerate}
\item[1.] $\overline{\pi}$ and $\pi^{\star \star}$ are both $^*$-representations of $A$, and $\pi \le \overline{\pi} \le \pi^{\star \star} \le \pi^*$. Moreover, one has that $\overline{D}=\bigcap_{x \in A} \overline{D}_x$.
\item[2.] $\pi$ is self-adjoint if and only if $D^* \subseteq D$.
\item[3.] $\pi^*$ is self-adjoint if and only if it is a $^*$-representation.
\item[4.] If $\pi$ is self-adjoint then any $^*$-representation extending $\pi$ in the same Hilbert space equals it.
\end{enumerate}
\end{prop}

Among those representations that usually appear in quantum statistical mechanics and quantum field theories, cyclic ones play a par\-ti\-cu\-lar\-ly relevant role. When dealing with algebras of non-necessarily bounded operators two definitions of cyclicity are available. 

Let $\pi$ be a representation of an algebra $A$ on a Hilbert space $\mathcal{H}$ and let $D$ be its domain. A vector $\xi \in D$ is said to be a {\it cyclic vector} if the set $\pi(A)\xi=\{\pi(x) \, \xi : x \in A\}$ is dense in $\mathcal{H}$. In that case we will say that the representation is a {\it cyclic representation}. If $\pi(A)\xi$ is dense in $D$ when endowed with the graph topology, then $\xi$ is said to be a {\it strongly cyclic vector} of $\pi$. A representation having a strongly cyclic vector will be called a {\it strongly cyclic representation}.

Let $\pi$ be a $^*$-representation of a $^*$-algebra $A$ on a Hilbert space $\mathcal{H}$ and let $\xi \in D$. Let $\hat{\pi}$ be the restriction of $\pi$ to $\pi(A)\xi$. It follows that $\xi$ is strongly cyclic if and only if the closure of $\hat{\pi}$ is an extension of $\pi$.

From now on, we will consider the case in which $A$ is a topological $^*$-algebra.

Let $\rho$ be a functional on $A$\footnote{Recall that we are calling functionals those elements in $A^*$.}. If for all $x \in A$ such a functional satisfies $\rho\left(x^*x\right) \geq 0$ then it will be called a {\it positive functional}. Continuous positive functionals over $A$ conform a proper strictly convex weakly closed cone in $A^*$ that we will denote $A^*_+$. While the sum and the multiplication law by non-negative real numbers are the ones induced by restricting the ordinary sum and scalar multiplication on $A^*$, the order on $A^*_+$ is defined as follows: given $\rho_1,\rho_2\in A^*_+$, one writes $\rho_1 \le \rho_2$ if and only if $\rho_1\left(x^*x\right) \le \rho_2\left(x^*x\right)$ for all $x\in A$. 

Let us recall that $A$ being a unital $^*$-algebra and $\rho$ being a positive functional on $A$, for all $x,y \in A$, it follows \cite{palmer2} that $\rho(x^*y)=\rho(y^*x)^*$. In particular, $\rho$ is {\it hermitian}, i.e., $\rho(x^*)=\rho(x)^*$ for all $x\in A$. Moreover, for all $x,y \in A$ it follows that $|\rho(x^*y)|^2 \le \rho(x^*x)\rho(y^*y)$ and $\rho$ is {\it Hilbert bounded}, i.e., there exists a constant $B$ satisfying, for all $x\in A$, $|\rho(x)|^2 \le B \rho(x^*x)$. The Hilbert bound $\|\rho\|\equiv \sup \{|\rho(x)|^2: x\in A, \rho(x^*x)\le 1\}$ equals $\rho(e)$, where $e$ is the unit in $A$.

We will say that a continuous positive functional on $A$ is an {\it extremal element of $A^*_+$} if it is indecomposable as a sum of continuous positive functionals that are not multiples of $\rho$. Equivalently, a continuous positive functional on $A$ is extremal in $A^*_+$ if and only if the interval $[0,\rho]=\{\rho'\in A^*_+:0\le \rho'\le\rho\}$ equals $\{\lambda \rho: 0\le\lambda\le 1\}$.

Finally, we can state the GNS theorem for topological $^*$-algebras. Its proof mainly follows the steps of the original proof for C$^*$-algebras (see, f.e., \cite{bratelli}). More details for the case of general $^*$-algebras can be found in \cite{powers}, \cite{palmer2} and \cite{schmudgen}. Notice that they all discuss pre-${^*}$-representations and only consider ${^*}$-representations for the normable case. This fact distinguishes our version of the theorem from theirs.

\begin{teo}
\label{teo2}
For each continuous positive functional $\rho$ on a topological $^*$-algebra $A$ with unit there is a closed weakly continuous strongly cyclic $^*$-re\-pre\-sen\-ta\-tion $\pi$ of $A$ on a Hilbert space $\mathcal{H}$ with domain $D$ such that
\begin{equation}
\label{normal}
\rho(x)=\left( \pi(x)\xi ,\xi\right)
\end{equation}
for all $x \in A$, $\xi \in D$ being a strongly cyclic vector of $\pi$. The representation $\pi$ is determined by $\rho$ up to unitary equivalence. Furthermore, if $\rho$ is an extremal functional then the corresponding representation is topologically irreducible.
\end{teo}

\begin{proof} Let $\rho$ be a continuous positive functional on $A$ and consider the quotient space $A_{\rho}=A/N_{\rho}$, where $N_{\rho}=\{x\in A: \rho(y^*x)=0 \mbox{ for all } y\in A\}$ is the so-called {\it Gelfand ideal} of $\rho$. Let us denote by $\phi_{\rho}$ the canonical projection of $A$ onto $A_{\rho}$ and let $(\cdot,\cdot):A_{\rho} \times A_{\rho} \rightarrow \mathbb{C}$ be the form on $A_{\rho}$ defined by $(\phi_{\rho}x,\phi_{\rho}y)=\rho(y^*x)$ for all $x,y\in A$. It is straightforward to check that this form is a positive non-degenerate sesquilinear form on $A_{\rho}$ endowing it with a pre-Hilbert structure. We will denote by $\mathcal{H}$ the completion of $A_{\rho}$ with respect to the norm $\|\cdot\|=(\cdot,\cdot)^{1/2}$.

Since $N_{\rho}$ is a left ideal of $A$, the map $\pi_0$ assigning to every element $x \in A$ the (non-necessarily bounded) operator $\pi_0(x)$ on $\mathcal{H}$ with domain $A_{\rho}$ given by $\pi_0(x)\phi_{\rho}y=\phi_{\rho}(xy)$ for all $y\in A$ defines a representation of $A$. It is actually a $^*$-representation of $A$ (see \cite{belanger} for the details). The closure $\pi$ of $\pi_0$ is the closed $^*$-representation of $A$ whose existence is claimed in the theorem. Recall that the domain $D$ of $\pi$ is the completion of $A_{\rho}$ with respect to the graph topology. The weak continuity of $\pi$ is a direct consequence of the continuity of $\rho$. On the other hand, setting $\xi=\phi_{\rho}e$, it follows that $\pi(x)\xi=\pi_0(x)\phi_{\rho}e=\phi_{\rho}x$ and, therefore, $\pi(A)\xi$ equals $A_{\rho}$, which is a dense subspace of $D$, showing that $\pi$ is a strongly cyclic representation of $A$ and that $\xi$ is a strongly cyclic vector of $\pi$. Finally, for all $x\in A$ we have that $(\pi(x)\xi,\xi)=(\pi_0(x)\phi_{\rho}e,\phi_{\rho}e)=(\phi_{\rho}x,\phi_{\rho}e)=\rho(x)$, and the first statement of the theorem is proved.

In order to prove that $\pi$ is determined by $\rho$ up to unitary equivalence it is necessary to show that there exists a unitary operator intertwining any two $^*$-representations satisfying (\ref{normal}). Explicitly, let $\pi_1$ and $\pi_2$ be two closed weakly continuous strongly cyclic $^*$-representations of $A$, let $\mathcal{H}_1$ (resp., $\mathcal{H}_2$) be the Hilbert space on which $\pi_1$ (resp., $\pi_2$) acts, let $D_1$ (resp., $D_2$) its domain, and let $\xi_1 \in D_1$ (resp., $\xi_2 \in D_2$) be a strongly cyclic vector of $\pi_1$ (resp., $\pi_2$) such that $(\pi_1(x)\xi_1,\xi_1)_1=(\pi_2(x)\xi_2,\xi_2)_2$, where $(\cdot,\cdot)_1$ (resp., $(\cdot,\cdot)_2$) is the inner product on $\mathcal{H}_1$ (resp., $\mathcal{H}_2$). We need to show that there exists a unitary operator $U$ mapping $\mathcal{H}_1$ onto $\mathcal{H}_2$ such that, for all $x\in A$, $U\pi_1(x)=\pi_2(x)U$. In \cite{powers} it was proved that such an operator is obtained by extending the operator $U_0:D_1 \rightarrow D_2$ given by $U_0\pi_1(x)\xi_1=\pi_2(x)\xi_2$ for all $x\in A$, to an operator from $\mathcal{H}_1$ into $\mathcal{H}_2$.

The proof of the last statement of the theorem concerning extremal functionals on $A$ and topologically irreducible representations of $A$ can be found in \cite{schmudgen}.
\end{proof}

Given a continuous positive functional $\rho$ on $A$, the representation built as in the theorem is called the {\it GNS representation of $A$} associated with $\rho$. The vector $\xi$ in (\ref{normal}) is sometimes referred as a {\it normalizing vector} of $\pi$.

GNS representations are usually constructed from states instead of positive functionals. {\it States over $A$} are defined as those continuous functionals that are positive and, in addition, satisfy $\rho(e)=1$. The set of states is a weakly closed convex section of $A^*_+$. {\it Pure states} are defined in analogy with extremal positive functionals as those that are indecomposable as convex combinations of other states. In that context it is possible to prove a stronger statement than the last one in Theorem 3.4. In fact, if GNS representations are built upon states, the cyclic vector $\xi$ is necessarily a normal vector, i.e., $\left(\xi, \xi \right)=1$, and the space of pure states is in one-to-one correspondence with the collection of unitary equivalence classes of topologically irreducible GNS representations of $A$.

At this point it will be convenient to introduce the space $\cycl(A)$ of those pairs of the form $(\pi,\xi)$ where $\pi$ is a closed weakly continuous strongly cyclic $^*$-representation of $A$ and $\xi$ is a particular strongly cyclic vector of $\pi$. We will endow this space with an equivalence relation as follows: given $(\pi_1,\xi_1)$ and $(\pi_2,\xi_2)$ in $\cycl(A)$ we will say that they are {\it unitarily equivalent}, and we will denote it by $(\pi_1,\xi_1) \sim (\pi_2,\xi_2)$, if there exists a unitary operator $U$ intertwining $\pi_1$ and $\pi_2$, i.e., $\pi_1(x) U=U \pi_2(x)$ for all $x\in A$, and, in addition, we have that $\xi_1=U\xi_2$. Notice that this notion of unitary equivalence for cyclic $^*$-representations is stronger than the usual one where no requirements on the corresponding cyclic vectors are imposed.

The motivations for introducing the space $\cycl(A)$ and their unitary equivalence classes arise from both physical and mathematical interests. From a strictly physical point of view, there are cases (f.e., when a symmetry is spontaneously broken) where the usual notion of unitary equivalence of $^*$-representations is not sufficient to ensure a complete identification of two physical situations. In a more mathematical context, the space $\cycl(A)$ and their unitary equivalence classes are relevant in our discussion since the quotient $\cycl(A)/\sim$ is precisely the space on which the GNS mapping becomes a bijection as it is proved in the following proposition.

\begin{prop}
\label{prop3}
The map assigning to each continuous positive functional $\rho$ on a topological $^*$-algebra with unit $A$ the GNS representation of $A$ associated with $\rho$ defines, up to unitary equivalence, a one-to-one mapping from $A^*_+$ onto $\cycl(A)$.
\end{prop}

\begin{proof} Theorem 3.4 asserts that the GNS mapping actually maps, up to unitary equivalence, $A^*_+$ into $\cycl(A)$. Therefore, we only need to check that this map is, in fact, a bijection. Let $\rho_1$ and $\rho_2$ be two continuous positive functionals on $A$ and let us assume that the corresponding GNS representations $\pi_1$ and $\pi_2$ are unitarily equivalent as elements of $\cycl(A)$. Let us denote by $\mathcal{H}_1$ and $\mathcal{H}_2$ the Hilbert spaces on which $\pi_1$ and $\pi_2$ act, and let $\xi_1$ and $\xi_2$ be their respective strongly cyclic vectors. It follows that there exists a unitary operator $U$ mapping $\mathcal{H}_1$ onto $\mathcal{H}_2$ satisfying $U\pi_1(x)=\pi_2(x)U$ for all $x\in A$ and $U\xi_1=\xi_2$. But then, for all $x\in A$, $(\pi_1(x)\xi_1,\xi_1)_1=(U\pi_1(x)\xi_1,U\xi_1)_2=(\pi_2(x)U\xi_1,U\xi_1)_2=(\pi_2(x)\xi_2,\xi_2)_2$, i.e., $\rho_1=\rho_2$. On the other hand, given an arbitrary element $\pi$ in $\cycl(A)$ with strongly cyclic vector $\xi$, the GNS representation of $A$ associated with the positive functional $\rho=(\pi(\cdot)\xi,\xi)$ is unitarily equivalent to $\pi$, the proof being identical to the one of the unicity statement in Theorem 2.
\end{proof}

The previous proposition shows that the space of unitary equivalence classes of $\cycl(A)$ can be suitably endowed with a proper strictly convex cone structure, the one inherited from $A^*_+$ through the GNS map. In the next section we will explicitly define this structure after relating the GNS representations of a $^*$-algebra with some of the Hilbert subspaces embedded in its anti-dual space.

Throughout the rest of the paper we will identify the elements in $\cycl(A)$ with their respective canonical GNS representatives and we will omit any explicit reference to the cyclic vector associated with each representation in $\cycl(A)$. Therefore, instead of saying that $(\pi,\xi)$ belongs to the unitary equivalence class in $\cycl(A)$ corresponding to the GNS representation associated with the functional $\rho=(\pi(\cdot)\xi,\xi)$ we will simply say that $\pi$ is an element of $\cycl(A)$.
 
\section{GNS representations and invariant Hilbert subspaces}
\label{gnshilb}

In the previous sections we have recalled the theory of the Hilbert subspaces of a quasi-complete locally convex Hausdorff separable vector space $E$ and we have discussed the GNS construction for a topological unital $^*$-algebra $A$. In this section, and following an idea already suggested in \cite{belanger} and \cite{constantinescu}, we will establish a connection between both approaches showing that there exists a bijection between $\cycl(A)$ and a particular subcone of $\hilb(A^*)$, this statement being valid for a wide class of topological $^*$-algebras. This characterization of $\cycl(A)$ will further allow us to explicitly endow it with a cone structure in such a way that this bijection actually becomes a cone morphism.

Let us begin with some general remarks on the continuous representations of a topological $^*$-algebra over a vector space and their restrictions to its Hilbert subspaces. Let $\pi$ be a strongly continuous representation of $A$ on $E$, i.e., a separately continuous map $(x,\phi) \rightarrow \pi(x)\phi$ from $A\times E$ into $E$ such that $\pi(xy)\phi=\pi(x)\pi(y)\phi$ for all $x,y \in A$ and all $\phi \in E$. Let us denote by $\pi^*$ the {\it dual representation} of $\pi$, i.e., the representation of $A$ on $E^*$ defined by $\langle \pi^*(x) y | \phi \rangle = \langle y | \pi(x^*) \phi \rangle$ for all $x\in A$, all $y\in E^*$ and all $\phi \in E$. The representation $\pi^*$ should not be confused with the Hilbert adjoint representation we have defined in Section 3. As it is the case for $\pi$, $\pi^*$ is also a strongly continuous representation of $A$ whenever $E^*$ is equipped with the weak topology, as we will assume from now on.

Given a Hilbert subspace $\mathcal{H}$ of $E$ with reproducing operator $H$, we will say that $\mathcal{H}$ is {\it invariant under $\pi$} or {\it $\pi$-invariant} if $\pi(x)HE^*\subseteq HE^*$ for all $x\in A$. If $\pi(x)H = H \pi^*(x)$ for all $x\in A$ we will say that $\mathcal{H}$ is {\it $^*$-invariant under $\pi$} or {\it $\pi$-$^*$-invariant}. Obviously, any $\pi$-$^*$-invariant Hilbert subspace of $E$ is invariant under $\pi$.

The motivation for introducing invariant and $^*$-invariant Hilbert subspaces is the following. If $\mathcal{H}$ is a $\pi$-invariant Hilbert subspace of $E$, the restriction of $\pi$ to $HE^*$ defines a representation of $A$ on $\mathcal{H}$ in the sense of Powers \cite{powers}. This fact should be clear since $\mathcal{H}$ can be thought as the completion of $HE^*$ with respect to the norm given by $\|Hx\|=\langle x | H x \rangle^{1/2}$ for all $x\in E^*$, i.e., $HE^*$ is an invariant dense subspace of $\mathcal{H}$.

When $\mathcal{H}$ is a $\pi$-$^*$-invariant Hilbert subspace of $E$, $\pi$ defines by restriction to $HE^*$ a $^*$-representation on $\mathcal{H}$. In fact, for all $x \in A$ one has that $\pi(x^*)$ equals $\pi(x)^*$ on $HE^*$ (see \cite{belanger} for the details). In this case, not only  the restriction of $\pi$ to $HE^*$ defines a $^*$-representation but also its closure, which exists since every $^*$-representation is closable. The domain $D$ of this representation, that we will denote also by $\pi$ as for the representation acting on $E$, is the completion of $HE^*$ in the graph topology.

Let us assume that $A$ is a barreled\footnote{Recall that a {\it barreled algebra} is a topological algebra in which every barrel, i.e., every absorbing, convex, balanced and closed subset, is a zero neighborhood \cite{treves}. \medskip} dual-separable $^*$-algebra with unit\footnote{Notice that this conditions are fulfilled in the particular case in which $A$ is separable by itself and $\{0\}$ is the intersection of a numerable set of environments. \medskip}. Since the weak dual of any barreled space is necessarily a quasi-complete space, Schwartz's theory of Hilbert subspaces applies to $A^*$ and we can set $E=A^*$ in the previous discussion. Under this identification $E^*$ is isomorphic to $A$.

Further, let us consider the particular case in which $\pi$ is the dual representation of the left regular action of $A$ on itself, i.e., the representation of $A$ on $A^*$ defined through $\langle y|\pi(x) \phi \rangle=\langle x^*y| \phi \rangle$ for all $x,y \in A$ and all $\phi \in A^*$. We will denote by $\hilb_A(A^*)$ the subcone of $\hilb(A)$ composed by those Hilbert subspaces of $A^*$ that are $^*$-invariant under the $\pi$.

\begin{teo}
\label{teorema}
For every $\pi$-$^*$-invariant Hilbert subspace $\mathcal{H}$ of $A^*$ there exists a closed weakly continuous strongly cyclic $^*$-representation of $A$ acting on it, this representation being identifiable with the GNS representation of $A$ associated with the functional $\rho=He$ where, as before, $H$ is the reproducing operator of $\mathcal{H}$ and $e$ is the identity of $A$. The correspondence defined in this way is a bijection from $\hilb_A(A^*)$ into $\cycl(A)$.
\end{teo}

\begin{proof} Let us first check that in this situation the restriction of $\pi$ to $HA$ defines a representation whose closure is a weakly continuous strongly cyclic $^*$-representation of $A$ on $\mathcal{H}$. The weak continuity of $\pi$\footnote{Recall that we are using the same notation for denoting both the representation acting on $A^*$ and the closed one on $\mathcal{H}$ defined by restriction} follows immediately, via polarizability, from the continuity of $H$, the strong continuity of $\pi$ on $A^*$ and the quasi-completeness of $A^*$. The strong cyclicity of $\pi$ is a consequence of the existence of a unit in $A$. Setting $\xi=He$ it follows that $\pi(A)\xi$ equals $HA$, and since this space is a subspace of $D$ that is dense for the graph topology, it follows that $\xi$ is a strongly cyclic vector of $\pi$. On the other hand, we have that $\rho(x)=\langle x | He \rangle=(\pi(x)He,He)$ for all $x \in A$\footnote{In what follows we omit any reference to the inclusion $J$ of $\mathcal{H}$ into $A^*$. Therefore, instead of writing $(J^*x,J^*y)$ for all $x,y \in A$, we will write $(Hx,Hy)$.}. By virtue of Theorem 3.4 it follows that $\pi$ can be identified with the GNS representation of $A$ associated with $\rho=He$.

We need to prove that the mapping we have defined is a one-to-one correspondence between $\hilb_A(A^*)$ and $\cycl(A)$. First, consider two different $\pi$-$^*$-invariant Hilbert subspaces of $A^*$. Let us denote them by $\mathcal{H}_1$ and $\mathcal{H}_2$. Let $H_1$ and $H_2$ be their respective reproducing operators and let $\pi_1$ and $\pi_2$ be the corresponding representations of $A$. Since $\mathcal{H}_1$ does not equal $\mathcal{H}_2$, it follows that $\rho_1=H_1 e$ is a functional on $A$ different from $\rho_2=H_2 e$. If it were not the case, a contradiction is obtained from the cyclicity of both functionals and the fact that $\pi_1$ and $\pi_2$ are restrictions of the same representation over $A^*$. Finally, since $\pi_1$ and $\pi_2$ are the GNS representations of $A$ associated with $\rho_1$ and $\rho_2$, respectively, from the unicity statement in theorem 2 it follows that $\pi_1$ and $\pi_2$ cannot be simultaneously identified with the same GNS representation.

Let us finally check that given a closed weakly continuous strongly cyclic $^*$-representation $\overline{\pi}$ of $A$ on a a Hilbert space $\overline{\mathcal{H}}$ with domain $\overline{D}$ and cyclic vector $\overline{\xi}$ it is possible to construct a $\pi$-$^*$-invariant Hilbert subspace of $A^*$ in such a way that the corresponding representation is equivalent to $\overline{\pi}$. Consider the correspondence $x \rightarrow \overline{\pi}(x) \overline{\xi}$. Let us denote it by $T$. Since $\overline{\pi}$ is weakly continuous, $T$ is a continuous operator mapping $A$ into $\overline{\mathcal{H}}$. On the other hand, $\overline{\pi}$ is strongly cyclic and then $T$ is a dense range operator. It follows that $T^*$, the adjoint of $T$, is an injective continuous map from $\overline{\mathcal{H}}$ into $A^*$. Let $\mathcal{H}$ be the image of $\overline{\mathcal{H}}$ through $T^*$ with the transported Hilbert space structure. The operator $H=T^*T$ from $A$ into $A^*$ is a positive operator reproducing $\mathcal{H}$ in $A^*$. It is easy to see that $\mathcal{H}$ is actually a $\pi$-$^*$-invariant Hilbert subspace of $A^*$. In fact, for all $x,y \in A$ we have that $\langle y | \pi(x) H e\rangle = \langle x^* y | H e \rangle = ( T(x^* y),Te )= (\overline{\pi}(x^*)Ty,Te)$. But since by hypothesis $\overline{\pi}$ is a $^*$-representation, we have that $(\overline{\pi}(x^*)Ty,Te)=(Ty,\overline{\pi}(x)Te)=(Ty,Tx)= \langle y | Hx\rangle$, i.e., $\pi(x) H e=Hx$, and then, $\pi(x) H y=H(xy)$ for all $x,y\in A$. Finally, $(\overline{\pi}(x)\overline{\xi},\overline{\xi})=(Tx,Te)=\langle x|T^*Te\rangle=\langle x|He\rangle$ and, then, $\pi$ can be fully identified with $\overline{\pi}$.
\end{proof}

We have found that, for a barreled dual-separable $^*$-algebra with unit $A$, there exists a multiple bijection among the following spaces:
\begin{enumerate}
\item the set $\cycl(A)$ of unitary equivalence classes of GNS representations of $A$,
\item the space $A^*_+$ of positive continuous functionals on $A$,
\item the cone of $\hilb_A(A^*)$ of Hilbert subspaces of $A^*$ that are $^*$-invariant under the dual left regular action $\pi$ of $A$,
\item the subfamily of $L^+(A^*)$ of continuous $\pi$-$^*$-invariant positive operators mapping $A^*$ into $A$, and
\item the space of $\pi$-$^*$-invariant positive kernels on $A$.
\end{enumerate}

The natural cone structures on the last four listed spaces are compatible with the bijection connecting them. It is then customary to transport such a structure on $\cycl(A)$. We introduce the following definitions.

{\bf Addition law in $\cycl(A)$:} As we will see in the next paragraphs, the multiplication law by non-negative real numbers and the order can be easily defined on $\cycl(A)$ without any reference to the bijection connecting this space with any other of those listed above. It is not the case for the sum. In fact, in order to properly define the sum of GNS representations it is mandatory to embed the corresponding representation spaces in a common domain. Let $\pi_1$ and $\pi_2$ be two elements of $\cycl(A)$ and let $\mathcal{H}_1$ and $\mathcal{H}_2$ be the two Hilbert subspaces of $A^*$ on which they act. Recalling that the domain of $\pi_1$ (resp. $\pi_2$) is the completion in the graph topology of $H_1A$ (resp. $H_2A$) where $H_1$ (resp. $H_2$) is the reproducing operator of $\mathcal{H}_1$ (resp. $\mathcal{H}_2$), let us consider the representation on $\mathcal{H}_1+\mathcal{H}_2$ with domain $(H_1+H_2)A$ given by $(H_1+H_2)y \rightarrow \pi_1(x)H_1y+\pi_2(x)H_2y$ for all $x,y\in A$. This is actually a well defined representation on $\mathcal{H}_1+\mathcal{H}_2$ since for any other $y'\in A$ such that $H_1y'+H_2y'=H_1y+H_2y$ it follows that $\langle z | \pi_1(x)H_1y'+\pi_2(x)H_2y' \rangle = \langle x^*z | H_1y'+H_2y' \rangle = \langle x^*z | H_1y+H_2y \rangle = \langle z | \pi_1(x)H_1y+\pi_2(x)H_2y \rangle$ for all $z\in A$ and then $\pi_1(x)H_1y'+\pi_2(x)H_2y'$ equals $\pi_1(x)H_1y+\pi_2(x)H_2y$. Moreover, since $\pi_1$ and $\pi_2$ are both weakly continuous strongly cyclic $^*$-representations of $A$, it is also the case for it. We will define the {\it sum of $\pi_1$ and $\pi_2$} as the closure of this representation and we will denote it by $\pi_1 + \pi_2$. Of course, $\pi_1 + \pi_2$ can be identified with the GNS representation of $A$ associated with the positive functional $H_1e+H_2e$.

{\bf Multiplication by non-negative real numbers in $\cycl(A)$:} Given a closed weakly continuous strongly cyclic $^*$-representation $\pi$ of $A$ acting on the $\pi$-$^*$-invariant Hilbert subspace $\mathcal{H}$ of $A^*$ we will define the representation $\lambda \pi$ for every $\lambda>0$ as the one that acts on $\lambda \mathcal{H}$ and algebraically coincides with $\pi$. If $\xi$ is the strongly cyclic vector associated with $\pi$, we will set $\lambda \xi$ to be the corresponding cyclic vector of $\lambda \pi$. The action of $\mathbb{R}_{>0}$ on $\cycl(A)$ is extended to $\mathbb{R}_{\ge 0}$ by identifying the representation $0 \pi$ with the trivial representation of $A$. It is straightforward to check that $\lambda \pi$ is a weakly continuous strongly cyclic $^*$-representation identifiable with the GNS representation of $A$ associated with $\lambda H e$ where, as before, $H$ is the reproducing operator of $\mathcal{H}$.

Notice that even when we are not identifying $\pi$ and $\lambda \pi$ as GNS representations of $A$, $\lambda \pi$ is unitarily equivalent to $\pi$ in the usual sense for every $\lambda > 0$. Consequently, extremal elements in $\cycl(A)$, i.e., those GNS representations of $A$ obtained as in Theorem 4.1 from extremal $\pi$-$^*$-invariant Hilbert subspaces of $A^*$, are necessarily associated with topologically irreducible representations.

{\bf Order in $\cycl(A)$:} A partial order on $\cycl(A)$ compatible with the operations we have just defined on this space has been already mentioned in the previous section. We will write $\pi_2 \le \pi_1$, $\pi_1$ and $\pi_2$ being two elements in $\cycl(A)$, if and only if $\pi_1$ extends $\pi_2$ in the sense of Remark \ref{obs}.

Finally we can state the following proposition. The proof straightforwardly follows from the previous definitions.

\begin{prop}
\label{cone}
The canonical bijection between $\cycl(A)$ and $\hilb_A(A^*)$ is an isomorphism for the cone structures we have defined. Explicitly, let $\pi$, $\pi_1$ and $\pi_2$ be three weakly continuous strongly cyclic $^*$-re\-pre\-sen\-ta\-tions of $A$ and let $\mathcal{H}$, $\mathcal{H}_1$ and $\mathcal{H}_2$ be three elements of $\hilb_A(A^*)$. Let $\lambda$ be a non-negative real number. It follows that
\begin{enumerate}
\item $\pi=\pi_1+\pi_2$ if and only if $\mathcal{H}=\mathcal{H}_1+\mathcal{H}_2$,
\item $\pi$ equals $\lambda \pi_1$ if and only if $\mathcal{H}=\lambda \mathcal{H}_1$, and
\item$\pi_1 \le \pi_2$ if and only if $\mathcal{H}_1 \le \mathcal{H}_2$.
\end{enumerate}
\end{prop}

\section{Consequences of the isomorphism between the spaces $\cycl(A)$ and $\hilb_A(A^*)$}
\label{consequences}

As we have already mentioned, the cone structure defined on $\cycl(A)$ have many interesting consequences. In this section we will derive some of them. As before, $A$ will denote a barreled dual-separable $^*$-algebra with unit. Hilbert subspaces of $A^*$ that are $^*$-invariant under the left dual regular action of $A$ on $A^*$ will be simply refered as $^*$-invariant Hilbert subspaces of $A^*$.

\begin{prop}
\label{prop5}
Let $\pi$ and $\pi_1$ be two elements in $\cycl(A)$. $\pi_1$ belongs to $[0,\pi]$ if and only if there exists a representation $\pi_2$ in $\cycl(A)$ such that $\pi=\pi_1+\pi_2$. In that case, $\pi_2$ is unique. We will denote it by $\pi_2 = \pi-\pi_1$.
\end{prop}

\begin{proof}
Let $\mathcal{H}_1$ and $\mathcal{H}$ be the $^*$-invariant Hilbert subspaces of $A^*$ associated with $\pi_1$ and $\pi$, respectively. If there exists a representation $\pi_2 \in \cycl(A)$ such that $\pi=\pi_1+\pi_2$ it follows that $\mathcal{H}=\mathcal{H}_1+\mathcal{H}_2$, $\mathcal{H}_2$ being the $^*$-invariant Hilbert subspace of $A^*$ associated with $\pi_2$ and then, $\mathcal{H}_1 \le \mathcal{H}$, i.e., $\pi_1 \le \pi$. Conversely, if $\pi_1 \in [0,\pi]$, $\mathcal{H}_1 \le \mathcal{H}$ and this inequality extends to the reproducing operators, i.e., $H_1 \le H$, where $H$ (resp., $H_1$) is the reproducing operator of $\mathcal{H}$ (resp., $\mathcal{H}_1$). It follows that $H-H_1$ is a positive operator reproducing a $^*$-invariant Hilbert subspace of $A^*$, say $\mathcal{H}_2$, whose associated representation $\pi_2$ in $\cycl(A)$ satisfies $\pi=\pi_1+\pi_2$. The uniqueness of $\pi_2$ follows from the uniqueness of the operator $H-H_1$.
\end{proof}

\begin{prop}
\label{prop6}
Let $\pi_1, \pi_2 \in \cycl(A)$. Then, $\pi_1$ is an algebraic subrepresentation of $\pi_2$ if and only if there exists $\lambda>0$ such that $\pi_1 \in [0,\lambda \pi_2]$.
\end{prop}

\begin{proof}
If there exists $\lambda>0$ such that $\lambda \pi_2$ extends $\pi_1$ it straightforwardly follows that $\lambda \pi_2$ algebraically extends $\pi_1$ and the same is true for $\pi_2$ since it is unitarily equivalent, in the ordinary sense, to $\lambda \pi_2$. Let us assume that $\pi_1$ is an algebraic subrepresentation of $\pi_2$. Let $\mathcal{H}_1$ and $\mathcal{H}_2$ the $^*$-invariant Hilbert subspaces of $A^*$ associated with $\pi_1$ and $\pi_2$, respectively. Since the inclusion of $\mathcal{H}_1$ into $A^*$ is continuous, its graph is closed in $\mathcal{H}_1 \times A^*$ and it is also the case for its graph in $\mathcal{H}_1 \times \mathcal{H}_2$. It follows from the closed graph theorem that the inclusion of $\mathcal{H}_1$ into $\mathcal{H}_2$ is continuous, and $\sqrt{\lambda}$ denoting its norm, we finally obtain that $\mathcal{H}_1 \le \lambda \mathcal{H}_2$. Therefore, $\pi_1$ is in $[0,\lambda \pi_2]$ as we wanted to prove.
\end{proof}

\begin{prop}
\label{prop7}
Let $\pi_1$ and $\pi_2$ be two representations in $\cycl(A)$. It follows that $\pi_1$ and $\pi_2$ are mutually excluding, i.e., $\pi_1+\pi_2$ is unitarily equivalent to $\pi_1 \oplus \pi_2$ if and only if $\pi\le \pi_1$, $\pi\le\pi_2$ with $\pi \in \cycl(A)$ implies $\pi=0$.
\end{prop}

\begin{proof}
Let $\mathcal{H}_1$ and $\mathcal{H}_2$ be the $^*$-invariant Hilbert subspaces of $A^*$ associated with $\pi_1$ and $\pi_2$, respectively. If $\pi_1+\pi_2$ is unitarily equivalent to $\pi_1 \oplus \pi_2$ it follows that $\mathcal{H}_1\cap\mathcal{H}_2={\{0\}}$ and then, if there exists a $^*$-invariant Hilbert subspace $\mathcal{H}$ of $A^*$ such that $\mathcal{H}\le\mathcal{H}_1$ and $\mathcal{H}\le\mathcal{H}_2$, we have $\mathcal{H}=\{0\}$, i.e., for any $\pi \in \cycl(A)$ such that $\pi\le \pi_1$ and $\pi\le\pi_2$ it follows that $\pi$ is the trivial representation. Conversely, assume that for any $\pi \in \cycl(A)$ satisfying $\pi\le \pi_1$ and $\pi\le\pi_2$ one has $\pi=0$ and let us define on $\mathcal{H}_1\cap\mathcal{H}_2$ the form $(\xi,\chi)=(\xi,\chi)_{\mathcal{H}_1}+(\xi,\chi)_{\mathcal{H}_2}$. This form is a positive definite inner product making $\mathcal{H}_1\cap\mathcal{H}_2$ into a Hilbert subspace of $A^*$ \cite{schwartz}. Since it clearly is a $^*$-invariant form, he have that $\mathcal{H}_1\cap\mathcal{H}_2$ belongs to $\hilb_A(A^*)$. But $\mathcal{H}_1\cap\mathcal{H}_2\le \mathcal{H}_1$ and $\mathcal{H}_1\cap\mathcal{H}_2\le \mathcal{H}_2$, and then $\mathcal{H}_1\cap\mathcal{H}_2=\{0\}$. Consequently, $\pi_1+\pi_2$ is unitarily equivalent to $\pi_1 \oplus \pi_2$, as we wanted to prove.
\end{proof}

\begin{prop}
\label{prop9}
Let $\pi_1$ and $\pi$ be two elements of $\cycl(A)$. Then, $\pi_1$ is a subrepresentation of $\pi$ in the ordinary sense if and only if $\pi-\pi_1 \in \cycl(A)$ and $\pi_1$ and $\pi-\pi_1$ are mutually excluding. In that case, $\pi-\pi_1$ is also a subrepresentation of $\pi$ in the ordinary sense and $\pi$ is unitarily equivalent to $\pi_1 \oplus (\pi-\pi_1)$.
\end{prop}

\begin{proof}
Let us first suppose that $\pi_1$ is unitarily equivalent to a subrepresentation of $\pi$ in the ordinary sense. It follows that the $^*$-invariant Hilbert subspace $\mathcal{H}_1$ of $A^*$ associated with $\pi_1$ is a subspace of the one associated with $\pi$ with the induced Hilbert space structure. Let us consider the space $\mathcal{H}_1^{\perp}$ orthogonal to $\mathcal{H}_1$ in $\mathcal{H}$. It is also a $^*$-invariant Hilbert subspace of $A^*$ and it equals $\mathcal{H}-\mathcal{H}_1$. Accordingly, we have that $\pi-\pi_1$ belongs to $\cycl(A)$ and that it is unitarily equivalent to a subrepresentation of $\pi$ in the ordinary sense, since the Hilbert space structure of $\mathcal{H}_1^{\perp}$ is the one induced by $\mathcal{H}$. On the other hand, since $\mathcal{H}_1\cap\mathcal{H}_1^{\perp}=\{0\}$, it follows from the previous proposition that $\pi$ is unitarily equivalent to $\pi_1 \oplus (\pi-\pi_1)$.

Reciprocally, let us assume that $\pi-\pi_1$ belongs to $\cycl(A)$ and that $\pi_1$ and $\pi-\pi_1$ are mutually excluding. If we denote by $\mathcal{H}_2$ the $^*$-invariant Hilbert subspace of $A^*$ associated with $\pi-\pi_1$, then $\mathcal{H}_1\cap\mathcal{H}_2=\{0\}$, $\mathcal{H}=\mathcal{H}_1+\mathcal{H}_2$ and $\mathcal{H}_1 \subset \mathcal{H}$ with the induced Hilbert space structure, i.e., $\pi_1$ is a subrepresentation of $\pi$ in the ordinary sense as we wanted to prove.
\end{proof}

\begin{prop}
Let $(\pi_i)_{i\in I}$ be a decreasing filtering system of representations in $\cycl(A)$. It follows that $\pi=\inf \{ \pi_i : i\in I\}$ exists in $\cycl(A)$.
\end{prop}

\begin{proof}
$I$ is a right filtering set of indices such that, for $i,j \in I$, $i \le j$, we have that $\pi_i$ is a subrepresentation of $\pi_j$. Let $(\mathcal{H}_i)_{i\in I}$ be the $^*$-invariant Hilbert subspaces of $A^*$ associated with $(\pi_i)_{i\in I}$, respectively, and let $(H_i)_{i\in I}$ be their corresponding reproducing operators. The space $\mathcal{H}=\inf \{ \mathcal{H}_i : i\in I\}$ exists in $\hilb(A^*)$ and its reproducing operator is $H=\inf \{H_i : i\in I\}$ which equals $\lim_i H_i$ in $L(A^*)$ when this space is endowed with the weak uniform convergence topology \cite{schwartz}. Since for every $i\in I$, $\mathcal{H}_i$ is $^*$-invariant under the dual left regular action of $A$, it follows that $\mathcal{H} \in \hilb_A(A^*)$. The corresponding GNS representation is the one whose existence is claimed in the proposition. Further, notice that the Hilbert space on which $\pi$ acts is the subspace of $\cap_{i\in I} \mathcal{H}_i$ composed by those $\phi  \in A^*$ such that $\|\phi\|:=\sup\{\|\phi\|_i:i\in I\}=\lim_i \|\phi\|_i< +\infty$.
\end{proof}

\begin{prop}
\label{propo}
Let $(\pi_i)_{i\in I}$ be an increasing filtering system of elements of $\cycl(A)$ and let $(\xi_i)_{i\in I}$ be their corresponding normalizing vectors. Then, $(\pi_i)_{i\in I}$ is majorized in $\cycl(A)$ if and only if
\begin{equation}
\label{condition}
\sup \{\|\pi_i(x)\xi_i\|^2_{i}:i\in I\}<+\infty
\end{equation}
for all $x\in A$. 
\end{prop}

\begin{proof}
Here, the set $I$ is as in the previous proposition but now for $i,j \in I$, $i \le j$, we have that $\pi_i$ is an extension of $\pi_j$. Let $(\mathcal{H}_i)_{i\in I}$ be the $^*$-invariant Hilbert subspaces of $A^*$ associated with $(\pi_i)_{i\in I}$, respectively, $(H_i)_{i\in I}$ being their reproducing operators. Notice that $\|\pi_i(x)\xi_i\|_i^2=\langle x, H_i x\rangle$ for all $i\in I$ and all $x\in A$. It follows that the condition (\ref{condition}) is necessary and sufficient for $(\mathcal{H}_i)_{i\in I}$ to be majorized in $\hilb_A(A^*)$. In this case, $\mathcal{H}=\sup\{\mathcal{H}_i:i\in I\}$ has $H=\sup\{H_i:i\in I\} = \lim_i H_i$ as a reproducing operator where, as before, the limit is taken in $L(A^*)$ with this space endowed with the weak uniform convergence topology. Since $\mathcal{H}$ is obviously $^*$-invariant under the dual left regular action of $A$, the proposition is proved. Notice, in addition, that $\mathcal{H}$ can be characterized as the $q$-completion in $A^*$ of the space $\cup_{i\in I} \mathcal{H}_i$ equiped with the pre-Hilbert structure derived from the norm $\|\phi\|:=\inf\{\|\phi\|_i:i\in I\}=\lim_i \|\phi\|_i$.
\end{proof}

\begin{prop}
Let $(\pi_i)_{i\in I}$ be a collection of elements of $\cycl(A)$. For every $i\in I$, let us denote by $\mathcal{H}_i$ the $^*$-invariant Hilbert subspace of $A^*$ associated with $\pi_i$, respectively. Consider the (abstract) Hilbert sum $\hat{\oplus}_{i\in I} \mathcal{H}_i$, the elements of this space being those sequences $(\phi_i)_{i\in I}$ with $\phi_i \in \mathcal{H}_i$ such that
\begin{equation}
\|(\phi_i)_{i\in I}\|^2=\sum_{i\in I} \|\phi_i\|_i^2 < +\infty
\end{equation}
where $\|\cdot\|_i$ denotes the norm in $\mathcal{H}_i$. Finally, let $\oplus_{i\in I} \mathcal{H}_i$ be the dense linear subspace of $\hat{\oplus}_{i\in I} \mathcal{H}_i$ composed by those sequences $(\phi_i)_{i\in I}$ in which all the $\phi_i$ are nul but a finite number of them and the pre-Hilbert structure inherited from $\hat{\oplus}_{i\in I} \mathcal{H}_i$.
It follows that the sums $\sum_{i\in I'} \pi_i$, $I'$ denoting the finite subsets of $I$, are majorized in $\cycl(A)$ if and only if the application $\Phi$ mapping $\oplus_{i\in I} \mathcal{H}_i$ into $A^*$ defined by
\begin{equation}
\label{condition33}
\Phi\left( \oplus_{i\in I} \phi_i \right)= \sum_{i\in I}\phi_i
\end{equation}
is continuous.
\end{prop}

\begin{proof}
Let us assume that the finite sums $\sum_{i\in I'} \pi_i$, $I'$ being any finite subsets of $I$, are majorized in $\cycl(A)$. From Proposition \ref{propo} we have that $\sum_{i\in I} \langle x|H_ix\rangle < +\infty$ for all $x\in A$, where we are denoting by $H_i$ the reproducing operator of $\mathcal{H}_i$ for all $i\in I$, respectively. In order to see that the mapping given by Eq. (\ref{condition33}) is continuous we must prove that the image of the unit ball $B$ in $\hat{\oplus}_{i\in I} \mathcal{H}_i$ under $\Phi$ is weakly bounded. Now, let $x$ be an element of $A$ and let $\phi$ be in $\Phi(B)$. $\phi$ equals $\sum_{i\in I'} \phi_i$ for a finite subset $I'$ in $I$ and $\sum_{i\in I'} \|\phi_i\|_i^2 < + \infty$. It follows that $|\langle x|\phi\rangle|^2 \le \sum_{i\in I'}|\langle x|\phi_i\rangle|^2 \le \left( \sum_{i\in I'} \|\phi_i\|_i^2 \right) \left( \sum_{i\in I'} \langle x|H_ix\rangle\right)^ \le \sum_{i\in I'} \langle x|H_ix\rangle$, what proves that $\Phi(B)$ is weakly bounded. Reciprocally, let us assume that $\Phi$ is a continuous mapping. Let us consider the extension of it to a continuous operator mapping $\hat{\oplus}_{i\in I} \mathcal{H}_i$ into $A^*$. If we denote by $\hat{\Phi}$ such extension and $(\phi_i)_{i\in I}$ is an element of $\hat{\oplus}_{i\in I} \mathcal{H}_i$, it follows that it equals the limit following the filtering system of finite subsets $I'$ of $I$ of those sequences $(\phi_i')_{i\in I}$ whose components satisfy $\phi_i'=\phi_i$ for all $i\in I'$ and $\phi_i'=0$ otherwise. Consequently, $\hat{\Phi}\left((\phi_i)_{i\in I} \right)$ is the limit taken in $A^*$ of $\sum_{i\in I'} \phi_i$. Further, we have the following factorization for $\hat{\Phi}$: $\hat{\oplus}_{i\in I} \mathcal{H}_i \rightarrow \left( \hat{\oplus}_{i\in I} \mathcal{H}_i \right)/\ker{(\hat{\Phi})} \rightarrow A^*$. The first mapping is the canonical projection of $\hat{\oplus}_{i\in I} \mathcal{H}_i$ onto $\left( \hat{\oplus}_{i\in I} \mathcal{H}_i \right)/\ker{(\hat{\Phi})}$, while the second one, that we will denote by $\tilde{\Phi}$, is an isomorphism from $\left( \hat{\oplus}_{i\in I} \mathcal{H}_i \right)/\ker{(\hat{\Phi})}$ onto the image of $\hat{\oplus}_{i\in I} \mathcal{H}_i$ under $\hat{\Phi}$. Assuming $\left( \hat{\oplus}_{i\in I} \mathcal{H}_i \right)/\ker{(\hat{\Phi})}$ endowed with the Hilbert structure derived from the quotient norm, let us consider on $\Phi\left( \hat{\oplus}_{i\in I} \mathcal{H}_i \right)$ this structure transported by $\tilde{\Phi}$, i.e., the metric structure making $\tilde{\Phi}$ an isometric isomorphism. Explicitly, the norm on $\Phi\left( \hat{\oplus}_{i\in I} \mathcal{H}_i \right)$ is given by $\|\phi\|_I^2=\inf\{\sum_{i\in I}\|\phi_i\|_i^2:\sum_{i\in I} \phi_i=\phi\}$. If $I'$ is a finite subset of $I$ it follows that any element $\phi$ in $\sum_{i\in I'}\mathcal{H}_i$ pick ups the form $\sum_{i\in I'}\phi_i$ with $\phi_i \in \mathcal{H}_i$ for all $i\in I'$, and then $\|\phi\|_{I'}^2=\inf\{\sum_{i\in I'}\|\phi_i\|_i^2:\sum_{i\in I'} \phi_i=\phi\}$. This shows that $\sum_{i\in I'}\mathcal{H}_i$ is a Hilbert subspace of $\Phi\left( \hat{\oplus}_{i\in I} \mathcal{H}_i \right)$. Further, since $\Phi\left( \hat{\oplus}_{i\in I} \mathcal{H}_i \right)$ is clearly a $^*$-invariant Hilbert subspace of $A^*$, it follows that the finite sums of the form $\sum_{i\in I'}\pi_i$ are majorized by the element in $\cycl(A^*)$ associated with $\Phi\left( \hat{\oplus}_{i\in I} \mathcal{H}_i \right)$.
\end{proof}

If a given sequence $(\pi_i)_{i\in I}$ of elements of $\cycl(A)$ satisfies the conditions of the previous proposition, it is called a {\it summable sequence} in $\cycl(A)$. The element in $\cycl(A^*)$ associated with $\Phi\left( \hat{\oplus}_{i\in I} \mathcal{H}_i \right)$ with the Hilbert structure transported by $\tilde{\Phi}$ is the {\it sum of $(\pi_i)_{i\in I}$} and we will denote it by $\sum_{i\in I}\pi_i$.

\begin{coro}
Let $(\pi_i)_{i\in I}$ be a collection of elements of $\cycl(A)$. Then, there exists a representation $\pi \in \cycl(A)$ such that $\pi = \oplus_{i \in I} \pi_i$ if and only if
\begin{enumerate}
\item[1. ] $\sum_{i\in I} \pi_i$ is well defined, and
\item[2. ] if $\phi_i \in  \mathcal{H}_i$ such that $\sum_{i\in I} \|\phi_i\|_i^2 < + \infty$ then $\sum_{i\in I} \pi_i(x)\phi_i=0$ for all $x\in A$ implies $\phi_i=0$ for all $i\in I$.
\end{enumerate}
\end{coro}

\begin{proof}
For the existence of such a representation it is necessary and sufficient that $(\pi_i)_{i\in I}$ is a summable sequence in $\cycl(A)$, i.e., $\sum_{i\in I} \pi_i$ is well defined, and that the map $\hat{\Phi}$ in the previous proposition is an isomorphism, i.e., the second condition.
\end{proof}

The definition of infinite sums of representations in $\cycl(A)$ has a natural generalization to integrals when $A$ by itself is weakly separable. We will briefly comment on this issue.

Let $\Gamma$ be a locally compact measure space and let us denote by $\mu$ its measure. We will say the a mapping  $\gamma \rightarrow \pi_{\gamma}$ from $\Gamma$ into $\cycl(A)$ is {\it integrable} if, for every $x\in A$, the function $\gamma \rightarrow \|\pi_{\gamma}(x)\xi_{\gamma}\|_{\gamma}^2$ is integrable, where we are denoting by $\xi_{\gamma}$ the normalizing vector of $\pi_{\gamma}$, for every $\gamma \in \Gamma$. The mapping $\gamma \rightarrow \mathcal{H}_{\gamma}$, where $\mathcal{H}_{\gamma}$ is the $^*$-invariant Hilbert subspace of $A^*$ associated with $\pi_{\gamma}$ for each $\gamma \in \Gamma$, respectively, will also be referred as an integral mapping from $\Gamma$ into $\hilb_A(A^*)$.

In \cite{schwartz} it was proved that given an integral map from $\Gamma$ into $\hilb_A(A^*)$ there exists a continuous mapping $\hat{\Phi}$ from $\int_{\Gamma}^{\oplus} \mathcal{H}_{\gamma} d\mu(\gamma)$ into $A^*$ defined by
\begin{equation}
\hat{\Phi}\left( (\phi_{\gamma})_{\gamma \in \Gamma}\right) =\int_{\Gamma} \phi_{\gamma} d\mu(\gamma)
\end{equation}
the second term in this equation being the weak integral of a scalarly integrable function. The space $\int_{\Gamma}^{\oplus} \mathcal{H}_{\gamma} d\mu(\gamma)$ is the space of measurable vector fields $\gamma \rightarrow \phi_{\gamma} \in \mathcal{H}_{\gamma}$ such that $\int_{\Gamma} \|\phi_{\gamma}\|^2_{\gamma} d\mu(\gamma) < +\infty$, where $\|\cdot\|_{\gamma}$ is the norm on $\mathcal{H}_{\gamma}$ for every $\gamma \in \Gamma$, endowed with the Hilbert space structure derived from the norm $\|(\phi_{\gamma})_{\gamma \in \Gamma}\|_{\Gamma}^2=\int_{\Gamma} \|\phi_{\gamma}\|^2_{\gamma} d\mu(\gamma)$.

As it is the case for infinite sums of Hilbert subspaces, the operator $\hat{\Phi}$ can be decomposed as $\int_{\Gamma}^{\oplus} \mathcal{H}_{\gamma} d\mu(\gamma) \rightarrow \left(\int_{\Gamma}^{\oplus} \mathcal{H}_{\gamma} d\mu(\gamma)\right)/\ker{(\hat{\Phi})} \rightarrow A^*$. The first map appearing in this factorization is the canonical projection of $\int_{\Gamma}^{\oplus} \mathcal{H}_{\gamma} d\mu(\gamma)$ onto $\left( \int_{\Gamma}^{\oplus} \mathcal{H}_{\gamma} d\mu(\gamma) \right)/\ker{(\hat{\Phi})}$ while the second one is the isomorphism from $\left(\int_{\Gamma}^{\oplus} \mathcal{H}_{\gamma} d\mu(\gamma)\right)/\ker{(\hat{\Phi})}$ onto $\hat{\Phi} \left(\int_{\Gamma}^{\oplus} \mathcal{H}_{\gamma} d\mu(\gamma)\right)$. This last space, if equipped with the transported Hilbert space structure of $\left( \int_{\Gamma}^{\oplus} \mathcal{H}_{\gamma} d\mu(\gamma) \right)/\ker{(\hat{\Phi})}$, is a Hilbert subspace of $A^*$ that we will denote by $\int_{\Gamma} \mathcal{H}_{\gamma} d\mu(\gamma)$. When all the vector fields $\gamma \rightarrow \phi_{\gamma}$ take values in $^*$-invariant Hilbert subspaces of $A^*$, i.e., when the integral mapping under consideration maps $\Gamma$ into $\hilb_A(A^*)$, the space $\int_{\Gamma} \mathcal{H}_{\gamma} d\mu(\gamma)$ turn to be also a $^*$-invariant Hilbert subspace of $A^*$, its associated representation in $\cycl(A)$ being called the {\it integral of the map $\gamma \rightarrow \pi_{\gamma}$}. Of course, we will denote it by $\int_{\Gamma} \pi_{\gamma} d\mu(\gamma)$.

Finally, we will discuss which is the effect of a continuous algebra $^*$-morphism in this context. Let $A_1$ and $A_2$ be two barreled dual-separable $^*$-algebras with unit and let $\alpha$ be a strongly continuous $^*$-homomorphism from $A_1$ into $A_2$. We will denote by $\pi_1$ (resp., $\pi_2$) the dual represntation of the left regular action of $A_1$ (resp., $A_2$) on its antidual space. We can prove the following proposition.

\begin{prop}
Let $\mathcal{H}_2$ be a $\pi_2$-$^*$-invariant Hilbert subspace of $A^*_2$ and let $\mathcal{H}_1$ be the Hilbert subspace of $A^*_1$ that is image of $\mathcal{H}_2$ under the transpose mapping of $\alpha$. It follows that $\mathcal{H}_1$ is $^*$-invariant under the dual left regular action of $A_1$.
\end{prop}

\begin{proof}
Let $H_1$ and $H_2$ be the reproducing operators of $\mathcal{H}_1$ and $\mathcal{H}_2$, respectively. Recall that $H_1=\alpha^* H_2 \alpha$. Let $x_1,y_1$ be an arbitrary pair of elements of $A_1$. Then $\langle y_1 | \alpha^* \pi_2(\alpha x_1) \phi_2 \rangle = \langle \alpha y_1 | \pi_2(\alpha x_1) \phi_2 \rangle = \langle (\alpha x_1)^* \alpha y_1 | \phi_2 \rangle = \langle \alpha (x_1^* y_1) | \phi_2 \rangle = \langle  x_1^* y_1 | \alpha^* \phi_2 \rangle = \langle  y_1 | \pi_1(x_1) \alpha^* \phi_2 \rangle$ , for all $\phi_2 \in A^*_2$, i.e., we have that $\alpha^* \pi_2(\alpha x_1)$ equals $\pi_1(x_1)\alpha^*$ on $A_2^*$. It follows that $\alpha^* \pi_2(\alpha x_1) H_2 \alpha = \pi_1(x_1) \alpha^* H_2 \alpha = \pi_1(x_1) H_1$. Now, let us assume that $\mathcal{H}_2$ is $^*$-invariant under $\pi_2$. Under this assumption we have $\alpha^* \pi_2(\alpha x_1) H_2 \alpha y_1 = \alpha^* H_2 [(\alpha x_1) \alpha y_1]=\alpha^* H_2 \alpha (x_1 y_1) = H_1(x_1 y_1)$. But then, for all $x_1,y_1 \in A_1$ we have that $\pi_1(x_1)H_1y_1=H_1(x_1 y_1)$, i.e., $\mathcal{H}_1=\alpha^*(\mathcal{H}_2)$ is $^*$-invariant under $\pi_1$, as we wanted to prove.
\end{proof}

The previous proposition shows that the mapping assigning to every Hilbert subspace $\mathcal{H}_2$ of $A_2^*$ the Hilbert subspace of $A^*_1$ given by $\mathcal{H}_1=\alpha^*(\mathcal{H}_2)$, when restricted to those Hilbert subspaces that are $^*$-invariant under the dual left regular action of $A_2$, defines a mapping from $\hilb_{A_2}(A_2^*)$ into $\hilb_{A_1}(A_1^*)$. Consequently, we have a well defined mapping from $\cycl(A_2)$ into $\cycl(A_1)$ that we will also denote by $\alpha^*$. It is straightforward to prove that this application is a cone structure preserving mapping.

\section{Conclusions and outlook}
\label{concluding}

Let us conclude this paper summarizing the main results we have obtained.

After recalling in section 2 the main aspects of Schwartz's theory on Hilbert subspaces of topological vector spaces, in section 3 we have discussed some basics of the representation theory of algebras of unbounded operators and we have restated the GNS construction theorem for general $^*$-algebras. In section 4 we have proved that for a wide class of topological $^*$-algebras, i.e., barreled dual-separable unital $^*$-algebras, their weakly continuous strongly cyclic $^*$-representations are in one-to-one correspondence with the Hilbert spaces continuously embedded in its dual that are $^*$-invariant under the dual left regular action of the algebra in hands. After explicitly endowing the first of these spaces with a cone structure we have proved that this correspondence actually is a cone isomorphism. Finally, in section 5 we have proved many consequences of the existence of such an isomorphism: we described the connection between the order of GNS representations and the usual concept of subrepresentation, we defined the difference of GNS representations, we proved a couple of propositions concerning the existence of extremal representations of filtering systems and we discussed the effect of $^*$-algebra morphisms.

As we have already mention these results could be useful for studying continuity aspects of the deformation of GNS representations. In a forthcoming paper we will prove that the similarity classes of GNS representations of a given barreled dual-separable $^*$-algebra with unit $A$ on inner product spaces are in one-to-one correspondence with the elements of the canonical real expansion of $L^+(A^*)$ showing that Kolmogorov functionals exhaustively define the GNS representations on Krein spaces.

\end{document}